# Relationships between distortions of inorganic framework and band gap of layered hybrid halide perovskites

Ekaterina I. Marchenko [1,2], Vadim V. Korolev [3], Sergey A. Fateev [1], Artem Mitrofanov [3,4], Nickolay N. Eremin [2], Eugene A. Goodilin [1,3], and Alexey B. Tarasov [1,3]*

[1] Laboratory of New Materials for Solar Energetics, Faculty of Materials Science, Lomonosov Moscow State University, 1 Lenin Hills, 119991, Moscow, Russia
[2] Department of Geology, Lomonosov Moscow State University, 1 Lenin Hills, 119991, Moscow, Russia
[3] Department of Chemistry, Lomonosov Moscow State University, 1 Lenin Hills, 119991, Moscow, Russia
[4] Science Data Software, LLC, 14909 Forest Landing Cir, Rockville, MD 20850, United States
* Author to whom correspondence should be addressed

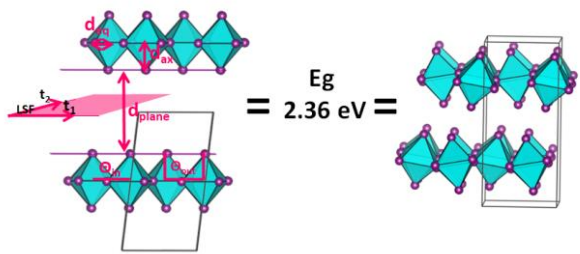

## Abstract

The unprecedented structural flexibility and diversity of inorganic frameworks of layered hybrid halide perovskites (LHHPs) rise up a wide range of useful optoelectronic properties thus predetermining the extraordinary high interest to this family of materials. Nevertheless, the influence of different types of distortions of their inorganic framework on key physical properties such as band gap has not yet been quantitatively identified. We provided a systematic study of the relationships between LHHPs' band gaps and six main structural descriptors of inorganic framework, including interlayer distances ($d_{int}$), in-plane and out-of-plane distortion angles in layers of octahedra ($\theta_{in}$, $\theta_{out}$), layer shift factor (LSF), axial and equatorial Pb-I bond distances ($d_{ax}$, $d_{eq}$). Using the set on the selected structural distortions we realized the inverse materials design based on multi-step DFT and machine learning approach to search LHHPs with target values of the band gap. The analysis of calculated descriptors – band gap dependences for the wide range of generated model structures of (100) single-layered LHHPs results in the following descending order of their importance: $d_{int} > \theta_{in} > d_{ax} > LSF_{min} > \theta_{out} > d_{eq} > LSF_{max}$, and also implies a strong interaction value for some pairs of structural descriptors. Moreover, we found that the structures with completely different distortions of inorganic framework can have similar band gap, as illustrated by a number of both experimental and model structures.

## Introduction

The family of layered hybrid halide perovskite-like compounds (LHHPs), often referred as "two-dimensional (2D) hybrid halide perovskites" is a new class of materials showing a set of unique functional properties such as record-breaking quantum yield of photo- and electroluminescence, tunable narrow or broad white-light emission, and excellent photoconductivity.[1–3] These properties of LHHPs are governed primarily by their crystal structure and therefore can be finely tuned in a broad range because of impressive structural flexibility and compositional diversity of this class of compounds[1,2,4]. LHHPs are formed as a result of the dimensionality reduction of the parent three-dimensional (3D) hybrid perovskites along specific crystallographic planes[5]. The general formula for the most common structural type with dimensionality reduction along (100) is $(A')_{2/q}A_{n-1}B_nX_{3n+1}$, where $[A']^{q+}$ is the singly (q = 1) or doubly (q = 2) charged organic interlayer (spacer) cation, $A^+$ is a small intralayer cation (such as $Cs^+$, $CH_3NH_3^+$, $[HC(NH_2)_2]^+$); $B^{2+} = Pb^{2+}$, $Sn^{2+}$, $Ge^{2+}$, etc.; $X^- = Cl^-$, $Br^-$, $I^-$; n is the number of layers consisting of the corner-shared octahedra within a perovskite slab.

A large number of suitable spacer cations has led to the rapid discovery of a long list of LHHPs, with over 600 of experimentally refined structures, well characterized by experimental methods and providing a valuable array of fundamental structural and optoelectronic data. Therefore, there is an emerging demand to apply computational and machine learning (ML) approaches[6] for the targeted search of new layered hybrid halide perovskites with desired properties.

Over the last years, many studies on crystal chemical analysis of layered hybrid perovskites have been published giving a better understanding of the structure-property relationships [7–12] and paving the way to predict new hybrid layered materials. The Density functional theory (DFT) has become the primary method for calculating the electronic properties of 3D hybrid perovskites. However, a much larger structural complexity of 2D HHPs in comparison with 3D ones hampers the possibilities of a wide-scale application of the common DFT-based approaches for the prediction of new LHHPs. Therefore, the combined approaches based on DFT and ML have been successfully applied to find the most promising structures in terms of band gap, effective mass of charge carriers, stability, and environmental friendliness (low toxicity)[13–17]. ML has helped dramatically to accelerate and to scale computational screening of LHHPs[6,18–20], thus discriminative models have already replaced partially the DFT calculations, since they link directly structural parameters of the materials with their properties[21–23]. However, revealing of the most reliable structure-property relationships of LHHPs is still complicated by a relatively few number of available experimental structures suitable for machine learning approaches.

In this study, we propose the multi-step approach based on ML and DFT for the rational search of 2D perovskites with any desired band gap based simply on the set of structural parameters describing the distortions of the inorganic framework and the interlayer distance. We investigated the role of each parameter and their mutual influence and contributions to achieve a target band gap.

**Methods**

In general, the proposed computational approach is based on so-called *inverse* materials design[24]. Within the direct material design, the parameters, such as atomic identities, composition, and structure *(ACS)*[24], often denoted as descriptors, act as inputs to the calculation algorithm, while the output is the target property *P*, such as a band gap or thermodynamic functions. Therefore, P being a function of ACS. In contrast, *inverse* materials design[24] uses the reverse sequence ("from *ACS* to *P(ACS)*" instead of "from *P(ACS) to ACS*") and sets the target property $P(ACS)$ as a starting point. Thus, the task boils down to an optimization problem in which the target property $P(ACS)$ acts as an objective function, and material descriptors $ACS$ form the search (i.e., design) space. This approach significantly accelerate the search, focuses on the desired property and allows to quantify the contribution of each of the initial parameters to the target property.

**Electronic structure calculations.** The band gap calculations using density functional theory (DFT) is a notorious problem. The most widely used semilocal exchange-correlation (XC) functionals[25] tend to underestimate the band gap value systematically[26]. Nevertheless, in three-dimensional perovskites, the values calculated using such XC functionals are close to the experimental ones due to cancellation errors caused by the opposite but similar in magnitude influence (~1 eV for lead perovskites[27]) of spin-orbit coupling (SOC) and many-body effects. However, this is not true for the two-dimensional perovskites, and the proper consideration of the many-body effects requires the use of post-DFT methods such as the GW approximation[28,29]. Unfortunately, this approach is not applicable for calculations in a high-throughput manner due to significant computational costs. One alternative that provides a speed-accuracy trade-off is the GLLB[30]-SC[31] potential, which in addition to the Kohn-Sham band gap $E_{gap}^{KS}$ directly incorporates the derivative discontinuity $\Delta_{xc}$. Thus, taking into account the SOC correction $\Delta_{SOC}$, the quasiparticle band gap $E_{gap}^{qp}$ considered in this work is calculated as follows:

$$E_{gap}^{qp} = E_{gap}^{KS} + \Delta_{xc} - \Delta_{SOC}$$

All presented DFT calculations were carried out with the projector-augmented wave (PAW) method[32] as implemented in the GPAW code[33,34]. Kohn-Sham wave functions[35] were expanded on a plane-wave basis with cut-off energy of 600 eV. The Brillouin zone was sampled using the Monkhorst-Pack algorithm[36]; the Γ-centered k-point mesh had a density of 5 k-points/Å. Fermi-Dirac smearing of 0.01 eV was also applied. The SOC correction was included in all the calculations.

Since the LHHPs are naturally represented as multiple-quantum-wells[37–39] formed by alternating organic and inorganic slabs, there are four types of band alignment[40] differing in the relative position of the lowest unoccupied /highest occupied molecular orbital (LUMO/HOMO) of the organic cation and conduction band maximum/valence band minimum (CBM/VBM) of the inorganic sheet. In practice, for the vast majority of known LHHPs, the type of band alignment is such that the HOMO and LUMO of cations have no considerable overlapping with VBM and CBM respectively[41–43]; both the low-energy electrons and holes are localized within the inorganic layer of octahedra. Accordingly, band gap energy is nearly fully determined by the configuration of the inorganic sublattice [44–48] whereas the nature and the position of organic cations affect the band gap indirectly, only through the distortion of inorganic layers[49]. In this case, the following trick works: molecular cations can be replaced by Cs atoms to calculate the band gap[47,50]. In order to verify the consistency of this approach, we calculated the band gaps for single-layered lead halide perovskites with aliphatic ammonium cations from a recently published database[11], for which the experimentally obtained values of the optical band gap are available. Organic moieties were replaced by Cs atoms, placed in the non-hydrogen atom position closest to the inorganic slab. The mean absolute deviation for a sample of 36 structures was 0.15 eV (Figure S1), which looks quite reasonable given the semilocal nature of XC functional used. Thereby, all the structures considered further, characterized by various distortions, were derived from the $Cs_2PbI_4$ prototype.

It should also be noted that the calculated values are systematically overestimated (the mean signed deviation was 0.11 eV). Apparently, this occurs due to the discrepancy between optical and quasiparticle band gaps. In particular, such omitting of the electron-hole interactions leads to a decrease in the band gap by only about 0.1 eV in the case of three-dimensional perovskites[51].

**Design space and descriptors.** The set of *descriptors* that compose the *design space* can be completely defined by the various distortions of the inorganic metal-halide framework. In particular, the following parameters were considered as structural descriptors influencing the band gap: axial and equatorial Pb-I distances, equatorial Pb-I-Pb and out of plane I-Pb-Pb-I angles (or I-Pb-Pb-I torsion angles), two components of layer shift factor (LSF)[12], and the interlayer distance. The presented octahedral tilting corresponds to $a^-a^-c^+$ type in Glazer notation[52], i.e., the combination of anti-phase tilt along [110] axis and rotation along [001] axis. The issue of the *descriptors selection* is discussed in the following section. Scales and valid ranges for each parameter are presented in the Table 1.

**Table 1.** Selected ranges of structural descriptors forming the design space.

| structural descriptor | min value | max value |
| --- | --- | --- |
| axial Pb-I distance ($d_{ax}$) | 3.15 Å | 3.25 Å |
| equatorial Pb-I distance ($d_{eq}$) | 3.15 Å | 3.25 Å |
| equatorial Pb-I-Pb angle ($Q_{in}$) | 150° | 180° |
| out of plane I-Pb-Pb-I angle ($Q_{out}$) | 150° | 180° |
| min component of LSF ($LSF_{min}$) | 0.0 | 0.5 |
| max component of LSF ($LSF_{max}$) | 0.0 | 0.5 |
| interlayer distance ($d_{int}$) | 4 Å | 7 Å |

**Optimization problem.** Inverse materials design is aimed at searching for materials with specific functionality[53]. In other words, if we consider a targeted property as a function of material's structure and we aim at finding a material with a record low value of this property, then the inverse design is reduced to an optimization problem[54-55]:

$$x^* = \underset{x \in \mathbb{X}}{\operatorname{argmin}} f(x)$$

where $\mathbb{X}$ is a design space (in our case — the set of values for six main structural descriptors –, $f(x)$ is an objective function (band gap), and $x^*$ is an optimal solution (set of structural descriptors' values corresponding to minimum band gap value). We can represent our task as a minimization problem since the desired band gap for the use in a single-junction (~1.3 eV[56]) or tandem (~1.75 eV[57]) solar cells lies below the range of values for experimentally obtained single-layer lead perovskites.

In this study, we used the sequential model-based optimization (SMBO) approach to solve the above optimization problem. Based on the set of accumulated observations $\mathcal{H}$ in the form of ordered pairs $(x_i, f(x_i))_{i=1}^N$, the posterior distribution over the initial objective function is computed using the Gaussian process[58] (GP). The acquisition function, which incorporates the posterior estimation, acquires its maximum value at the next point $x_{i+1}$ to be evaluated. Here we used the GP-Hedge algorithm[59] to sample points nominated with three acquisition functions: lower confidence bound, negative expected improvement, and negative probability of improvement. The mean and covariance functions implemented to define GP were a constant function and Matérn kernel, respectively. The optimization procedure was implemented using the scikit-optimize package. In the initial stage, 20 observations were randomly sampled from the design space (Table 1). Then, 30 SMBO steps were performed using a posterior estimation at each iteration.

**Interpretation of structure-property relationships.** The SMBO approach does not *per se* provide insights into the structure-property relationships. To define the relevance of each structural descriptor for the band gap, we used the surrogate-model feature attribution method, namely the Shapley additive explanations (SHAP) approach[60]. Shapley values[61] from cooperative game theory[62] provide a quantitative estimation of the importance of individual observations and satisfy three important properties (local accuracy, consistency, and missingness). More particularly, we used their specific version, SHAP values[63,64], which were calculated using the TreeExplainer algorithm[60] designed for tree-based discriminative models. SHAP interaction values[65] were implemented to capture local interaction effects determined by the interplay between descriptors. The SHAP analysis was provided using the gradient boosting algorithm implemented in the XGBoost library[66]. An ensemble of 1000 models was used to neglect the effect of train-test data splitting. Visualization of crystal structures was carried out using the VESTA program[67].

To analyse the interaction of organic cations in such structures we studied two-dimensional fingerprint plots derived from the Hirshfeld surface of a given cation using the Crystal Explorer [68]. These plots represent a fraction of points on the Hirshfeld surface as a function of the closest distances from the point to atom inside ($d_i$) and outside ($d_e$) the surface and so indicate not only what kind of interactions are present, but also the relative area of the surface corresponds each to a particular interaction. For example, the two distances, $d_e$ and $d_i$, are defined as the distances from the point on the surface to the nearest exterior atom in the neighboring cation and, respectively, to the closest interior atom in the same cation. It is straightforward to map on the Hirshfeld surface and elucidate the differences in packing of interlayer organic cations between different polymorphs.

**Results and discussion**

The band structure of (100) single-layered hybrid halide perovskites is mainly determined by the overlapping between metal and halogen orbitals within the layers as well as between p-orbitals of halogen atoms of different layers in the case of small interlayer distances which makes the band gap value to be mainly affected by the geometry of the inorganic framework[50]. The vast majority of organic cations (except ones with expanded π-conjugated systems) do not per se affect the electronic density of states near VBM and CBM influencing the band gap only indirectly through the distortion of inorganic framework. Without the distortions of metal-halide layers of octahedrons all single-layered perovskites would have the same value of band gap, which contradicts the experimentally observed large scatter of their optical band gap values. Therefore, it is reasonable to use the inorganic framework distortions as main structural descriptors linked with the band gap of LHHPs. To select the particular set of structural descriptors, we analyzed the recent studies on the influence of various structural parameters on the band gap of LHHPs.

Optical studies and DFT calculations showed that for lead halide hybrid layered perovskites with butylammonium cations and various intralayer organic cations (methylammonium, formamidinium, dimethylammonium and guanidinium) the elongation of Pb–I bond length reduces the overlap of the Pb s- and I p-orbitals and increases the optical bandgap [48]. Similarly, increasing of the tilting angles of octahedra quantified by I-Pb-Pb-I out of plane angles ($Q_{out}$) also reduces the Pb-I overlapping and, therefore, increases the bandgap[46,50]. In addition, the band gap has been found to increase with the increase of the interlayer distance (d) in LHHPs[69]: it was shown that the $E_g(d)$ increases mainly within the range of 3.5–4.5 Å and further reaches a plateau at d > 4.6 Å for the $A_2PbBr_4$ hypothetical compound. Such a dependency of Eg(d) originates obviously from the contribution of a relatively weak interlayer electronic coupling of the orbitals of terminal halogen atoms.[69] Finally, the layer shift factor (LSF) between adjacent layers of octahedra affects the distortion of the inorganic framework and the band gap values:

the calculated band gap values of hypothetical (100)-single-layered lead bromide perovskites with interlayer distances of 3.5 Å monotonically increase with the increase in the LSF values from (0, 0) to (1/2, 1/2)[12].
Consequently, to characterize the mutual arrangement of Pb and I ions in LHHPs, we chose the following set of six structural descriptors: Pb-I axial and equatorial bond distances ($d_{ax}$ and $d_{eq}$ respectively); in-plane distortion Pb-I-Pb ($\theta_{in}$) and out-of-plane distortion I-Pb-Pb-I ($\theta_{out}$) angles in layers of octahedra; the interlayer distances between planes of iodine ions ($d_{plane}$) in adjacent layers of octahedra; layer shift factor - LSF($t_1$, $t_2$). Notably, we used the interlayer distances between planes of axial iodine ions ($d_{plane}$) instead of the commonly used minimal distances between iodine ions ($d_{int}$) because for the distorted structures $d_{plane} < d_{int}$, whereas $d_{plane} = d_{int}$ only for LSF (0.0, 0.0).
To examine the influence of structural parameters to the band gap of (100) single-layered LHHPs we generated a set of structures with different distortion descriptors and applied Bayesian optimization scheme to find the structure with a minimum band gap value, as shown in Figure 1a. As expected, these structures were found to be undistorted and have the following descriptors values: $\theta_{in}=\theta_{out}=180$, $d_{ax}=d_{eq}=3.15$ Å (corresponds to the average Pb-I bonds in 3D lead halide perovskite), LSF(0,0), $d_{int}$ = 4 Å. To analyze the influence of distortion parameters we use the individual SHAP values, either positive or negative, on the band gap (Figure 1b), because traditional variable importance algorithms only show the results across the entire population but not on each individual case. As seen from Figure 1b, the most important descriptor was found to be the interlayer distance ($d_{int}$) and all the descriptors can be ranged in the following descending order of importance: $d_{int} > \theta_{in} > d_{ax} > LSF_{min} > \theta_{out} > d_{eq} > LSF_{max}$. The value of the descriptor is indicated with a colored gradient where blue and red correspond to minimum and maximum values of the descriptors taken from the chosen range (see Table 1). The broadening of the lines at a given SHAP value is proportional to the number of structures with a corresponding impact on the band gap value and different combinations of other descriptors. In general, the decrease of the $d_{int}$, $d_{ax}$, and $d_{eq}$ values decrease the band gap, whereas the decrease of the $\theta_{in}$ and $\theta_{out}$ values increases it.
Further we analyze the mutual influence of each pair of the distortion parameters based on SHAP interaction values. The combination of $d_{int}$ and $\theta_{in}$ were found to be the most influencing on the band gap (for $d_{int}$ = 4 – 7 Å), while ($\theta_{out}$, $d_{ax}$) and ($LSF_{max}$, $d_{ax}$) combinations are next in importance (Figure 1c). The LSF ($t_1$,$t_2$) and $d_{in}$ parameters influences the band gap value mostly when $d_{in}$ distances are shorter than 6 Å[69]. Importantly, if we use the distance between the planes formed by the halogens of neighboring layers ($d_{plane}$) instead of the minimum distance between halogen atoms from adjacent layers ($d_{int}$), it turns out to be much less convenient for optimization. As shown by the SHAP analysis, the minimum allowable value of $d_{plane}$ affects the band gap to a lesser extent than $d_{int}$ (Figure 2 a-b). The difference in the influence between $d_{plane}$ and $d_{int}$ on the band gap is in a full agreement with the physical meaning of these parameters (the distance between two nearest iodine atoms of adjacent layers and distance between two geometrical planes of axial iodine atoms respectively).

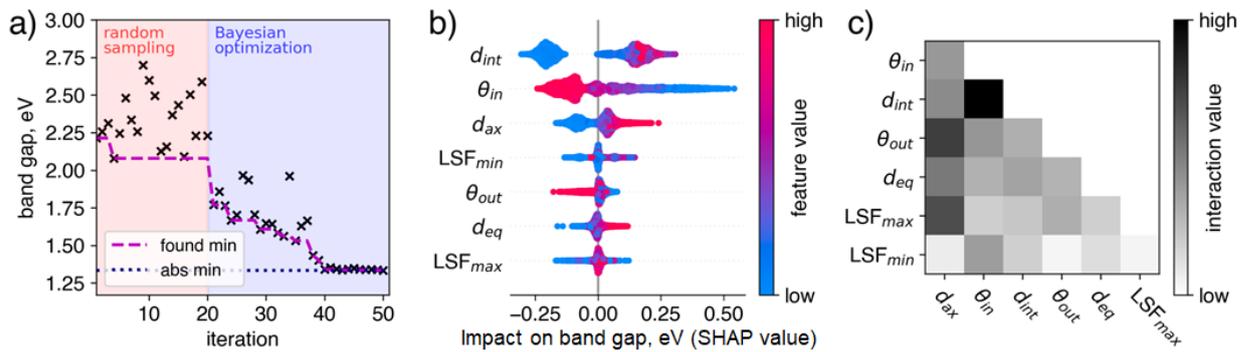

Figure 1. Iterative search for LHHPs with a minimum band gap (a), SHAP values (b) and SHAP interaction values (c) with $d_{int}$ descriptor as interlayer distances between layers of octahedra.

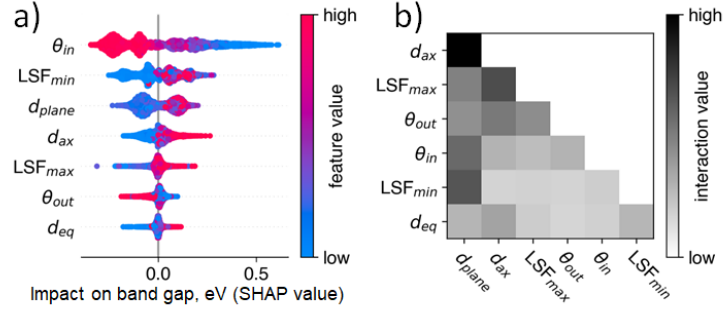

Figure 2. SHAP values (a) and SHAP interaction values (b) with $d_{plane}$ parameter as interlayer distances between layers of octahedra.

Then we analyze the variation of the band gap with two distinct structural descriptors for the model structures of single-layered iodoplumbate LHHPs with 3 different interlayer distances (4, 5 and 6 Å) and fixed values of other parameters. In particular, the following pairs of descriptors were considered: (a) I-Pb-Pb-I out of plane angle – equatorial Pb-I-Pb angle, (b) axial Pb-I distance – equatorial Pb-I distance, (c) minimum and maximum translation components of LSF – $t_1$ and $t_2$. The band gap functional dependencies were plotted in a form of a 2D color map for each pair with a given value of $d_{int}$ (Figure 3). First of all, it is worth noting that all of these features are smooth and monotone, which reflects the symbatic nature of the influence of each of the descriptors in each pair on the band gap energy ($E_g$). The second important consequence is that the contribution of internal layer parameters, $\theta_{in}$ and $\theta_{out}$ angles (Figure 3a) and $d_{ax}$ and $d_{eq}$ distances (Figure 3b), to the band gap value remain almost the same for all tree interlayer distances, whereas the contribution of $LSF_{max}$ and $LSF_{min}$, being the interlayer parameters, dramatically reduces and becomes virtually negligible with an increase from 4 to 6 Å (Figure 3c). These results clearly illustrate that the contributions of some structural descriptors to the band gap value can vary significantly with altering other ones, especially for pairs of descriptors with a strong interaction value.

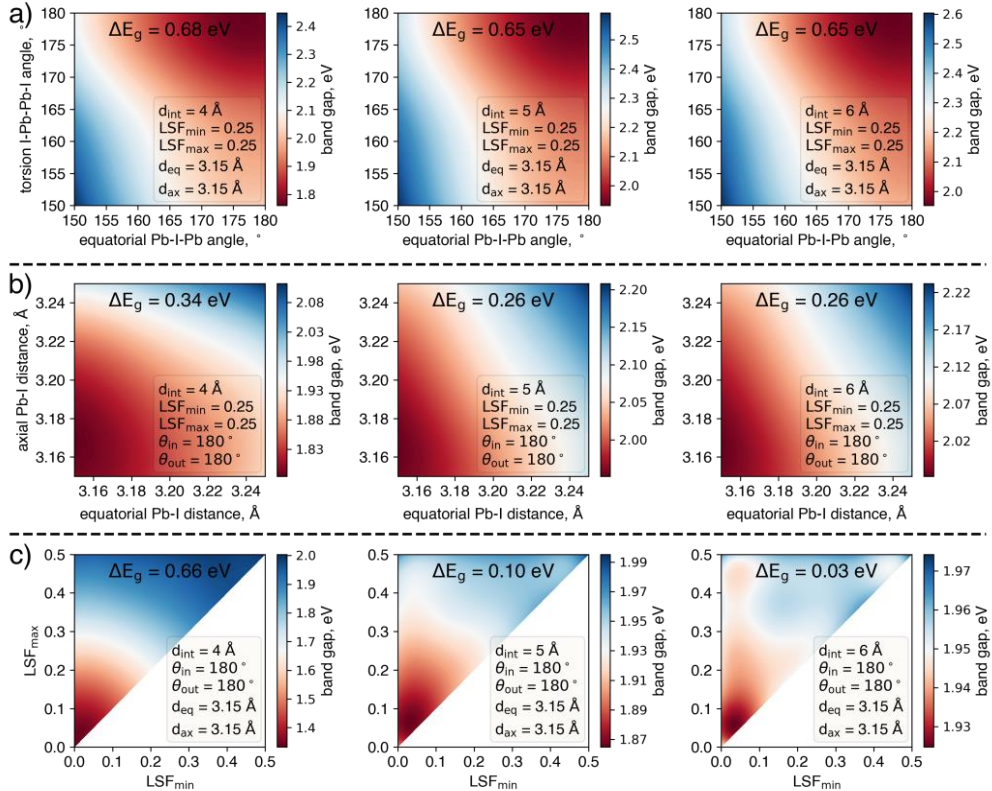

Figure 3. 2D color maps illustrating the variation of calculated band gaps with (a) $\theta_{out}$ angle – equatorial Pb-I-Pb angle, (b) axial Pb-I distance – equatorial Pb-I distance, and (c) minimum and maximum components of LSF ($t_1$, $t_2$)

for the series of model $Cs_2PbI_4$ single-layered (100) perovskites with the tree different interlayer distances (4, 5 and 6 Å).

Another important conclusion is that the simultaneous influence of structural distortions of inorganic frameworks on the band gap is such, that the same target value of band gap can also be achieved by different combinations of descriptors. For the further expansion of this statement we analyzed the database[11] of LHHPs experimentally refined structures and found the compounds with significantly different sets of structural descriptors but the same values of experimentally measured optical band gap.

Indeed, the structures ID 221[70] (with 2-ethyl-hexylammonium cation), ID 465[71] (with 1-(4-chlorophenyl)ethylammonium) and ID 474[72] (with imidazolium ethylammonium) exhibit significantly different inorganic framework distortions but demonstrate almost identical band gaps of 2.36 eV (Figure 4). Obviously, such different distortions of the inorganic layer of iodoplumbate octahedrons in these three structures are determined by the templating effect of three completely different organic interlayer cations. In addition to these experimental structures, we show that the structure *ID 13* from our set of modelled structures (see SI file) has a similar band gap but differing features of the inorganic lattice geometry. For the structures ID221, and ID465, the same value of the band gap is achieved by the superposition of $d_{ax}$, $\theta_{in}$ and $d_{eq}$ contributions, whereas LSF and the interlayer distance are not expected to be meaningful parameter because of negligible overlapping between the adjacent layers of $PbI_6$ octahedra for $d_{int} > 6$ Å. Unlike the previous ones, the structure ID474 has a small interlayer distances (3.667 Å) and therefore $d_{int}$ and LSF parameters significantly affect the band gap. Other examples of the experimentally investigated structures with the same values of the band gap (equal to 2.04 eV and 2.28 eV respectively) but different organic cations are represented by the ID 290[73] - ID 317[74] and ID 473[72] - ID 475[75] pairs (Figures S2 and S3 in SI). All these examples illustrate the possibility of mutual compensation between the different structural descriptors in term of effect on band gap.

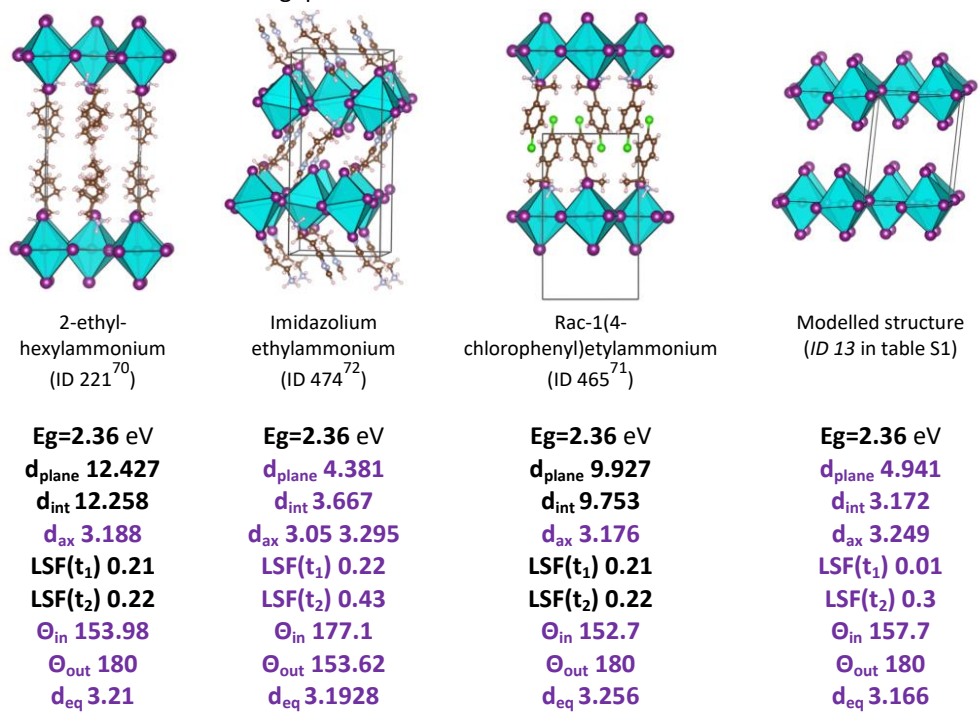

| 2-ethyl-hexylammonium (ID 221[70]) | Imidazolium ethylammonium (ID 474[72]) | Rac-1(4-chlorophenyl)etylammonium (ID 465[71]) | Modelled structure (*ID 13* in table S1) |
|---|---|---|---|
| **Eg=2.36** eV | **Eg=2.36** eV | **Eg=2.36** eV | **Eg=2.36** eV |
| $d_{plane}$ 12.427 | $d_{plane}$ 4.381 | $d_{plane}$ 9.927 | $d_{plane}$ 4.941 |
| $d_{int}$ 12.258 | $d_{int}$ 3.667 | $d_{int}$ 9.753 | $d_{int}$ 3.172 |
| $d_{ax}$ 3.188 | $d_{ax}$ 3.05 3.295 | $d_{ax}$ 3.176 | $d_{ax}$ 3.249 |
| $LSF(t_1)$ 0.21 | $LSF(t_1)$ 0.22 | $LSF(t_1)$ 0.21 | $LSF(t_1)$ 0.01 |
| $LSF(t_2)$ 0.22 | $LSF(t_2)$ 0.43 | $LSF(t_2)$ 0.22 | $LSF(t_2)$ 0.3 |
| $\Theta_{in}$ 153.98 | $\Theta_{in}$ 177.1 | $\Theta_{in}$ 152.7 | $\Theta_{in}$ 157.7 |
| $\Theta_{out}$ 180 | $\Theta_{out}$ 153.62 | $\Theta_{out}$ 180 | $\Theta_{out}$ 180 |
| $d_{eq}$ 3.21 | $d_{eq}$ 3.1928 | $d_{eq}$ 3.256 | $d_{eq}$ 3.166 |

Figure 4. Experimental (ID 221, 474, 465) and theoretically generated (*ID 13*) LHHPs structures with different interlayer sets of structural distortions and the same band gaps (2.36 eV).

As it was shown on the analysis of SHAP interaction values, the highest mutual influence of descriptors is observed for the case of structures with small interlayer distances. Although it is problematic to select such pairs from the set of experimental structures, we can distinguish them from the set of the structures generated for ML purpose (see SI file). Both pairs of selected structures are characterized by relatively small variations in interplane distances, whereas the first pair of such structures, ID28 and ID35, with band gap of 1.61 eV has significantly different axial Pb-I bond lengths and LSF values (3.15 and 3.25 Å, 0.39 and 0.31 respectively, Figure 5, a) and the structures of the second pair ID29 and ID30, with calculated band gap 1.64 eV differ only in equatorial Pb-I distance (Figure 5, b). Thus, we can conclude that in the case of structures with small interlayer distances, even a small change in the LSF and $d_{plane}$ can compensate for the effect on the band gap from a significant increase in the bond length, which is apparently explained by a large contribution of interlayer overlapping. This once again demonstrates the importance of simultaneous considering of all the selected structural descriptors for the structures with $d_{int} < 5$ Å.

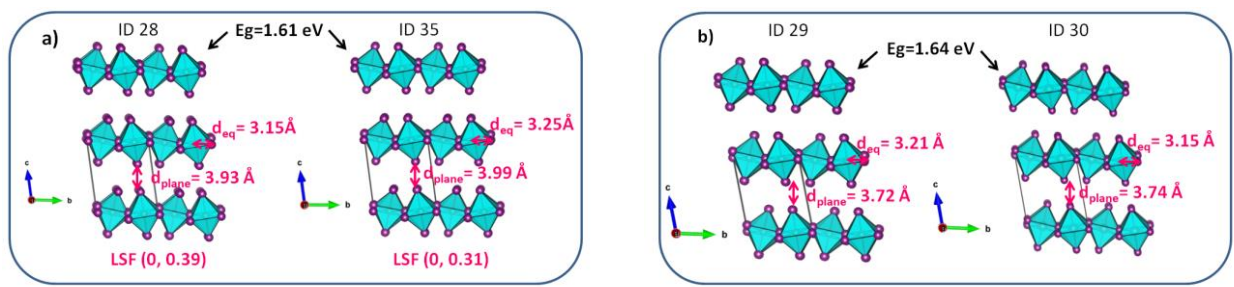

Figure 5. Two couples of generated structures with different descriptors and the same bang gap: a) $d_{eq}$, $d_{plane}$, LSF and $E_g$ = 1.61 eV; b) $d_{eq}$, $d_{plane}$ and $E_g$ = 1.64 eV.

As shown above, considering the distortions of inorganic Pb-I frameworks quantified in the form of selected structural descriptors makes it possible to effectively evaluate their effect on the band gap and, accordingly, to construct inorganic frameworks for new possible structures with the desired band gap values. However, to reach the final goal of the inverse material design of new LHHPs it is necessary to reveal relationships between the distortions of the inorganic framework and the packing of organic cations in the interlayer space. Useful model objects for this purpose are polymorphs, a change in the character and density of packing of cations in which with temperature leads to distortions of the inorganic perovskite layer.

Several compounds from the Database have polymorphs at different temperatures featured by the different packing of large organic cations. For example, the LHHPs with hexylammonium cations demonstrate a high temperature (at T=273 K) and low temperature (T=173 K) polymorphs characterized by different values of structural descriptors determined by different packing of organic cations. Particularly, the structures ID 20[76] and ID 21[76] have distinctly different two-dimensional fingerprint plots of organic cations (Figure 6). It is worth noting that minimal $d_e$ = 1.07 Å in ID 21 appear to be nonphysical values and are apparently related to the orientational disorder of the H atoms. A similar dependency towards a general increase in the volume of organic cations in polymorphic modifications of structures at different temperatures is observed for a number of experimentally known structures with chain cations: pentylammonium, hexylammonium, heptylammonium, octylammonium and nonylammonium (see Figures S4-S7 in SI).

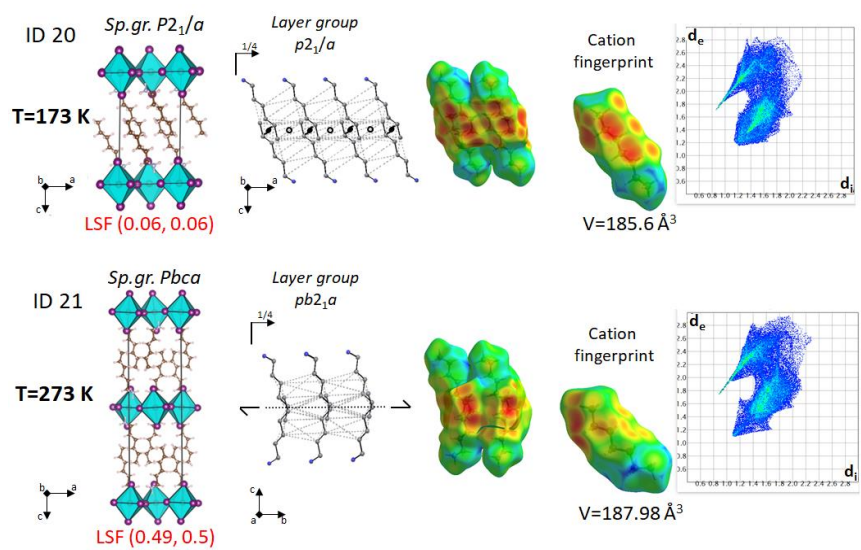

Figure 6. Polymorphic modifications of single-layered LHHPs with hexylammonium organic cations with different topology of organic layers at different temperatures.

All the structural descriptors in ID 20 and ID 21 are different, while the LSF parameter changes the most (from (0.06,0.06) for low temperature polymorph ID 20 to (0.49, 0.5) for higher temperature polymorph ID 21) accompanied by a difference in overlapping of orbitals. Interestingly, there is an unambiguous relationship between LSF and cation packing. Among the two polymorphs (ID 20 and ID 21), the denser packing of organic cations in layers in a higher temperature polymorph increases the symmetry of the organic layers from monoclinic ($p2_1/a$) to orthorhombic ($pb2_1a$) with the loss of an inversion center in the organic layer in favor of a denser packing (Figure 6). Thus, a denser packing of cations increases the symmetry of the inorganic layers from monoclinic to orthorhombic. However, a detailed exploration of this relationship between packing of different organic layers and structural descriptors is a subject for further research.

## CONCLUSIONS

To sum up, we present a first systematic study of the effect of the main structural parameters of LHHPs' inorganic framework on their band gaps. We use six structural descriptors (interlayer distances, in-plane and out-of-plane distortion angles in layers of octahedra, layer shift factor, axial and equatorial Pb-I bond distances) to construct the design space in order to realize the inverse material design approach to search LHHPs with target values of a band gap. The proposed multi-step DFT and ML based approach has shown its high efficiency of the searching for a band gap minimum and, therefore, for prediction of LHHPs with a given $E_g$.

The analysis of calculated descriptor – band gap dependences for the wide range of generated model structures of (100) single-layered LHHPs results in discovery of several new structure-property correlations. First, we range all structural descriptors in the following descending order of importance: $d_{int} > \theta_{in} > d_{ax} > LSF_{min} > \theta_{out} > d_{eq} > LSF_{max}$. Moreover, we revealed strong interaction value for some pairs of structural descriptors (the most important for $\theta_{out}, d_{ax}$ and $LSF_{max}, d_{ax}$ pairs), which means that they would not be considered as independent parameters but should be jointly used. For another pair of descriptors $\{d_{int}, LSF\}$ it was shown that the elongation of the former from 4 to 6 Å dramatically reduces and makes virtually negligible the contribution of the LSF to the resulting band gap. These results clearly illustrate that the contributions of a given structural descriptor to the band gap value can vary significantly with altering other ones, especially for pairs of descriptors with a strong interaction value.

Another important conclusion is that the simultaneous influence of structural distortions of inorganic framework on the band gap is such, that the same target value of band gap can also be achieved by a wide range of various combinations of descriptors, as illustrated by a number of both experimental and model structures.

## Supporting Information

Examples of generated crystal structures of (100)- lead halide perovskites with different distortion descriptors, calculated and experimental band gaps for (100)- single-layer lead halide hybrid structures examples of LHHPs, structures with different interlayer organic cations and the same band gaps, analysis of packing topology of organic "layers" of LHHPs polymorphs with the same organic cations.

## AUTHOR INFORMATION

### Corresponding Author
* alexey.bor.tarasov@yandex.ru

### Author Contributions
The manuscript was written through contributions of all authors. All authors have given approval to the final version of the manuscript.

### Notes
The authors declare no competing financial interest.

## ACKNOWLEDGMENT
This work was financial supported by a grant from the Russian Science Foundation, project number 19-73-30022.